\documentstyle[aps,amsfonts,pre,preprint,epsfig]{revtex}

\begin{document}

\renewcommand{\theequation}{\arabic{section}.\arabic{equation}}

\newcommand {\bv}[1]{ \mathbf{#1} } \newcommand {\mchi}{\bar{\chi}}
\newcommand {\hv}[1]{\hat{\bv #1}} \newcommand
{\gf}[2]{\Gamma\!\left(\frac{#1}{#2}\right)}

\author{ A.N. Morozov$^{1,2}$ \and A.V. Zvelindovsky$^2$ \and
J.G.E.M. Fraaije$^2$}

\address{$^1$Faculty of Mathematics and Natural Sciences, University
of Groningen, Nijenborgh 4, 9747 AG Groningen, The Netherlands, email:
a.n.morozov@chem.rug.nl}

\address{$^2$Soft Condensed Matter group, LIC, Leiden University, PO
Box 9502, 2300 RA Leiden, The Netherlands}

\title{Influence of confinement on the orientational phase transitions
in the lamellar phase of a block copolymer melt under shear flow}

\maketitle

\begin{abstract} In this work we incorporate some real-system effects
into the theory of orientational phase transitions under shear flow
(M. E. Cates and S. T. Milner, Phys. Rev. Lett. {\bf 62} 1856 (1989)
and G. H. Fredrickson, J. Rheol. {\bf 38}, 1045 (1994)). In
particular, we study the influence of the shear-cell boundaries on the
orientation of the lamellar phase. We predict that at low shear rates
the parallel orientation appears to be stable. We show that there is a
critical value of the shear rate at which the parallel orientation
loses its stability and the perpendicular one appears immediately
below the spinodal. We associate this transition with a crossover from
the fluctuation to the mean-field behaviour. At lower temperatures the
stability of the parallel orientation is restored. We find that the
region of stability of the perpendicular orientation rapidly decreases
as shear rate increases. This behaviour might be misinterpreted as an
additional perpendicular to parallel transition recently discussed in
literature.
\end{abstract}

\section{ Introduction }

When subjected to a shear flow, AB block copolymer melts exhibit an
orientational phase behaviour which is absent in equilibrium.  A
system under shear shows not only transitions between different
morphologies (typically lamellar, hexagonal, cubic and gyroid
\cite{hamley:book,Leibler}), but also transitions between different
{\it orientations} of these morphologies with respect to the shear
geometry. Experimental literature extensively discusses this effect
for lamellar \cite{koppi92,koppi93} and hexagonal phases
\cite{tepe,walter98}.

\begin{figure}
\begin{picture}(0,610)
\put(-50,0){\epsfig{file=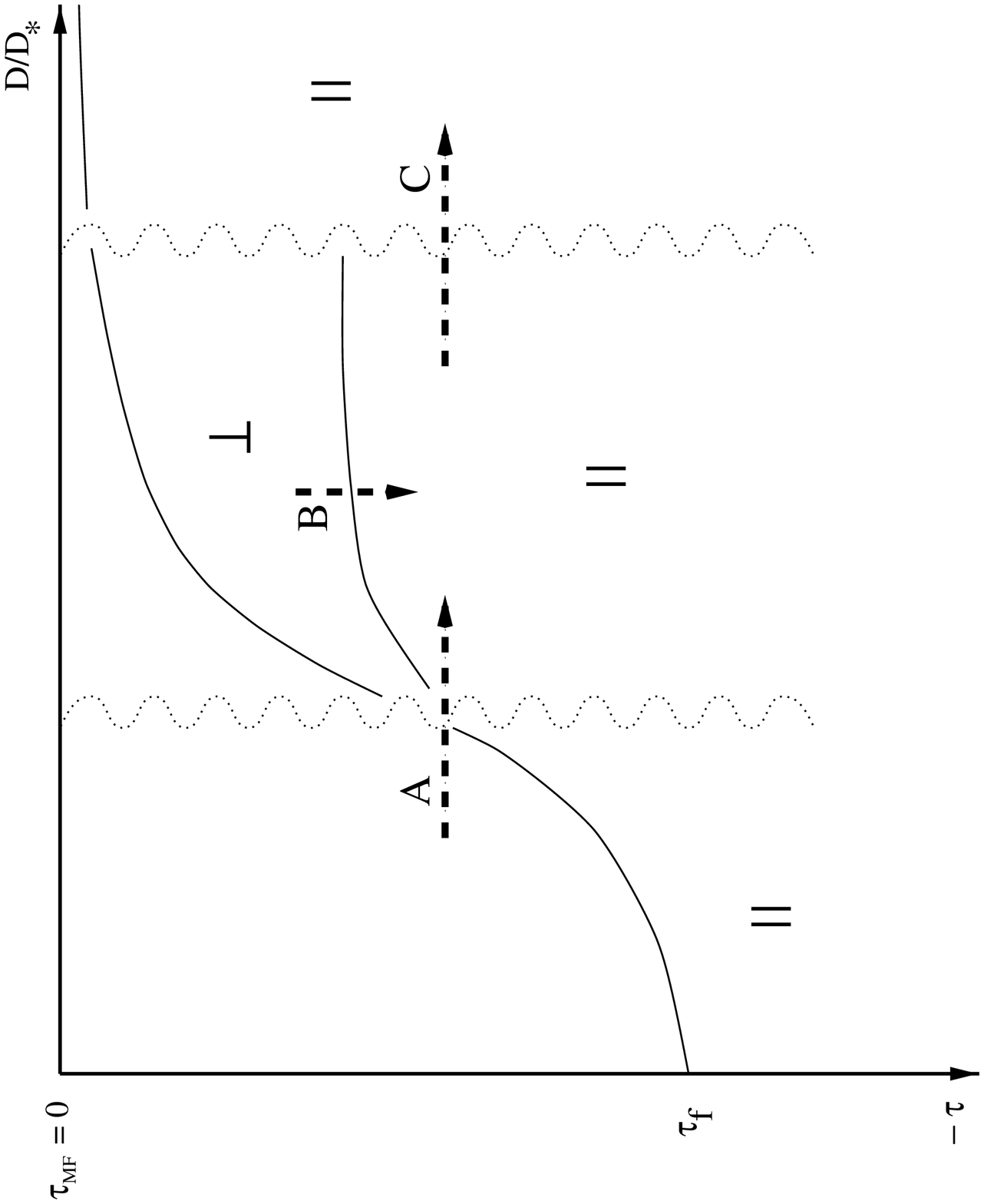,width=17cm}}
\end{picture}
\caption{}
\label{fig1}
\end{figure}

The theoretical description of the lamellar reorientation was
developed in \cite{cates_milner,j_rheol}. The same method was applied
in \cite{ik_hexag} to study the hexagonal pattern. In these theories
orientational transitions appear as a result of interaction of shear
flow with critical fluctuations in melt. There are two distinct
regimes: a slow flow only slightly perturbs the fluctuation spectrum
while a fast flow significantly dumps fluctuations, restoring the
mean-field behaviour in the limit of infinite shear rate
$D\rightarrow\infty$. Correspondingly, the parallel lamellae (their
normal is parallel to the shear gradient direction) are found to be
stable in the small shear rate regime, while the perpendicular
lamellae (their normal is perpendicular to both the gradient and flow
directions) are stable at high shear rates. Fredrickson has shown that
if one takes into account the difference in viscosities of the pure
melt components, the perpendicular phase loses its stability at low
enough temperatures and the parallel orientation is
restored. Schematically this behaviour is summarized in
Fig.\ref{fig1}.

However, there is an experimental evidence that this picture is not
complete. At very high shear rates the parallel orientation was found
to be the only stable one \cite{zhang95,patel95}. This cannot be
explained in the discussed framework of \cite{j_rheol}, since it
predicts the stability region for the perpendicular phase {\it to
increase} as $D\rightarrow\infty$.

In this work we propose an explanation of the additional transition
(C-transition in Fig.\ref{fig1}). We argue that the missing element of
the theory is the interaction of the block copolymer melt with the
walls of the shear cell. We consider a block copolymer film confined
in-between two walls in the gradient direction and subjected to a
steady shear flow. Usually the distance between the interfaces in the
other two directions is much larger and we ignore their influence.
This model will predict the parallel orientation to be stable in the
$D\rightarrow\infty$ limit since the influence of shear and
fluctuations vanishes in this limit. The only symmetry-breaking factor
is then the wall-copolymer interaction which stabilizes the parallel
orientation \cite{fr_surf,binder_films}. The complex behaviour at
lower shear rates will arise from the interplay of three factors:
shear flow, fluctuations and wall-melt interactions.

We admit that the influence of the surface interactions is possibly
small. However, Balsara {\it et. al} reported \cite{balsara94} that in
the absence of shear the walls of their shear cell induced the
parallel alignment through the whole $0.5$-mm--sample, although the
lamellar spacing is somewhat 4 orders of magnitude smaller. Under
shear Laurer {\it et. al} \cite{laurer} observed that independently of
the bulk orientation there is always a near-surface layer of the
parallel lamellae which penetrates up to $2\,\mu$m into the
bulk. Thus, even a weak symmetry-breaking field can be crucial in the
absence of other factors.

We also want to mention that the equilibrium theory of block-copolymer
melt ordering near surfaces is well-developed
\cite{fr_surf,binder_films,freed92,binder94,chakrabarti,chakrabarti94,binder97,matsen97,agur_film}.
Some questions about dynamics of such an ordering were addressed in
\cite{agur_dynamics,binder97}. However, until now this theory was
never applied to non-equilibrium systems.

Our paper is organized as follows. In Section II we derive the
equations governing the dynamics of the melt and construct a
non-equilibrium potential whose minimal value will determine the
stable orientation. In the first part of Section III we estimate the
shear rate of the A-transition while the other two transitions (B and
C) are analyzed in the second part. In conclusion we discuss in detail
properties of the obtained phase diagram.  In Appendix we provide an
example clarifying the role of thermal fluctuations.

\section{Dynamic equations} \setcounter{equation}{0}

\begin{figure}
\begin{picture}(0,610)
\put(0,0){\epsfig{file=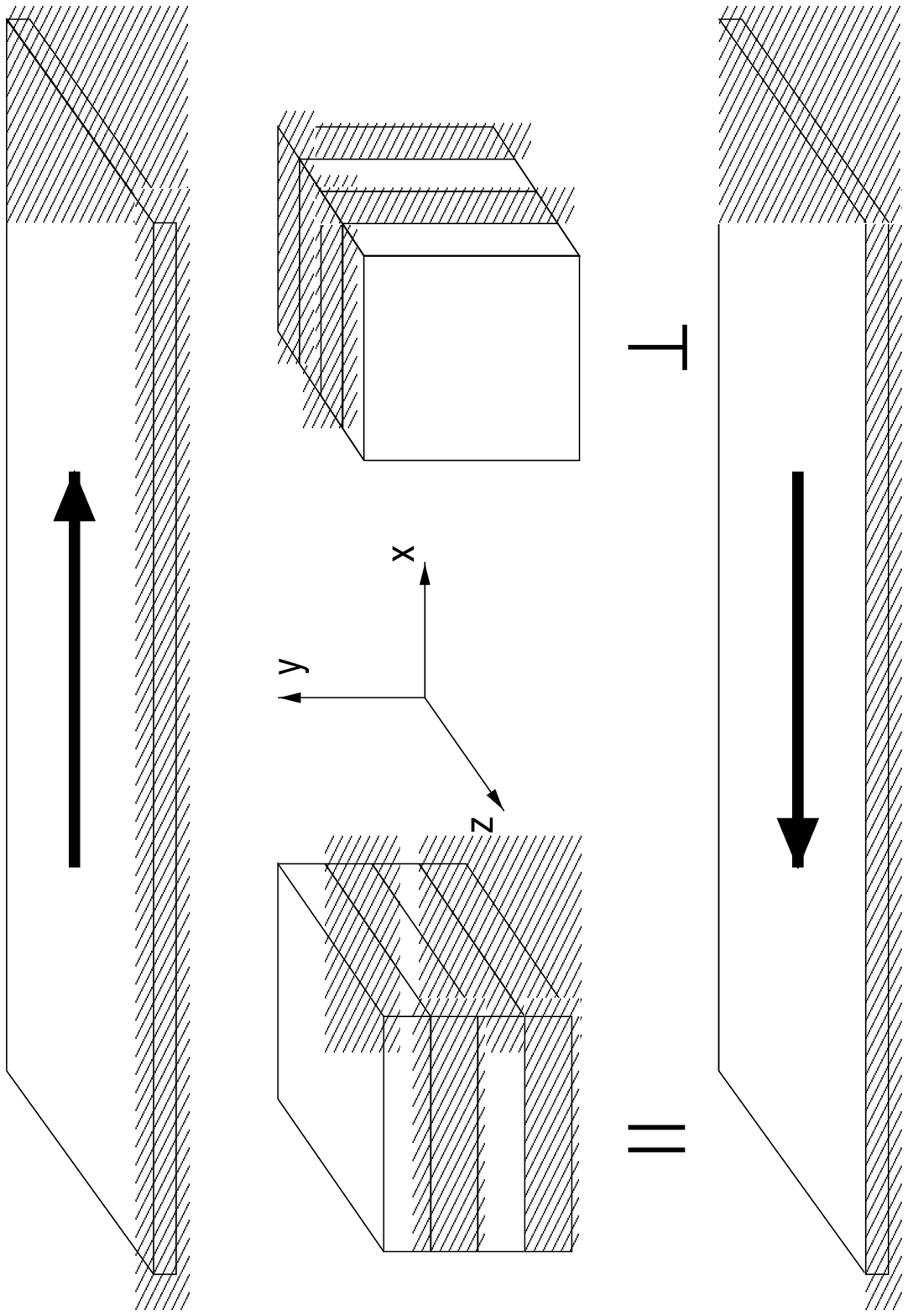,width=13cm}}
\end{picture}
\caption{}
\label{fig2}
\end{figure}

Let us consider a block copolymer melt confined in-between two
surfaces in the $y$-direction. It is also subjected to a steady shear
flow ${\bv v}=D y {\bv e}_x$ (see Fig.\ref{fig2}). We ignore any
alteration of this velocity profile and assume that it is kept through
the whole system. We choose the local deviation of composition from
its average to be an order parameter $\phi({\bv r})$ and define its
Fourier transform as
\begin{equation} \phi({\bv k})=\int \! d{\bv r}\,e^{-i {\bv k r}}
\phi({\bv r}) \qquad \text{and} \qquad \phi({\bv r})=\int_k e^{i {\bv
k r}} \phi({\bv k})
\end{equation} where
\begin{equation} \int_k \equiv \frac{1}{L}\sum_{k_y}\int \frac{d k_x d
k_z}{(2\pi)^2}\qquad \text{and} \qquad \int \! d{\bv r} \equiv
\int_{-\infty}^{\infty}dx \int_{-L/2}^{L/2}dy \int_{-\infty}^{\infty}
dz
\end{equation} It is convenient to work in dimensionless units and we
rescale lengths and wave-vectors: ${\bv r}\rightarrow b^{-1} {\bv r}$
and ${\bv k} \rightarrow b \,{\bv k}$, $b$ being the size of a
monomer.

Following \cite{onuki_kawasaki,cates_milner,j_rheol} we assume that
the dynamics under shear flow is governed by the Fokker-Planck
equation:
\begin{equation}
\label{FP} \frac{\partial P[\phi,t]}{\partial t}=\int\!\!d{\mathbf
{r}}\, \frac{\delta }{\delta
\phi({\mathbf{r}})}\left[\mu\left(\frac{\delta }{\delta
\phi({\mathbf{r}})}+\frac{\delta {\mathcal H}}{\delta
\phi({\mathbf{r}})} \right) + D y \frac{\partial \phi}{\partial
x}\right] P[{\mathbf {r}},t]
\end{equation} where $P[\phi,t]$ is the probability to realize the
order parameter profile $\phi({\bv r})$ at time $t$, and $\mu$ is an
Onsager coefficient. In Appendix we provide arguments for using the
Fokker-Planck equation instead of any other deterministic equation.

In eq.(\ref{FP}) the Hamiltonian ${\mathcal H}$ consist of two
contributions: the bulk Hamiltonian derived by Leibler \cite{Leibler}
\begin{eqnarray} N {\mathcal H}_L[\phi]&=&\frac{1}{2}\int_q
\Gamma_2(q)\phi({\bv q})\phi(-{\bv q})+\frac{1}{3!}\int_{q_1}
\int_{q_2} \int_{q_3} \Gamma_3({\bv q}_1,{\bv q}_2,{\bv q}_3)
\phi({\bv q}_1)\phi({\bv q}_2)\phi({\bv q}_3) \nonumber \\
&+&\frac{1}{4!} \int_{q_1} \int_{q_2} \int_{q_3} \int_{q_4}
\Gamma_4({\bv q}_1,{\bv q}_2,{\bv q}_3,{\bv q}_4) \phi({\bv
q}_1)\phi({\bv q}_2)\phi({\bv q}_3) \phi({\bv q}_4)
\end{eqnarray} and the surface energy \cite{fr_surf,binder_films}
\begin{equation}
\label{Hs} N {\mathcal H}_s=\int\!\!d{\mathbf {r}} \left[-H_1
\phi({\mathbf r})+\frac{a_1}{2}\phi({\mathbf r})^2
\right]\left[\delta\left(y+\frac{L}{2}\right)+\delta\left(y-\frac{L}{2}\right)\right]
\end{equation} where $N$ is a number of monomers in a molecule,
$H_1\sim (\chi N)_{cop-surf}$ is the strength of the interaction
between the surface and the copolymer melt and $a_1$ describes the
additional interaction in the melt induced by the presence of the
surface (it changes the local temperature in the vicinity of the
surface). Our goal is to construct a real-space version of the Leibler
Hamiltonian ${\mathcal H}_L$. In \cite{Ohta:1986,Cruz:1986,fr_surf} it
was shown how to deal with the second-order vertex
function. Separating small- and large-wave-vector asymptotic
behaviour, one can show that
\begin{equation} \Gamma_2(q)\approx\frac{A}{q^2}+B q^2 -\mchi
\end{equation} where
\begin{eqnarray}
\label{const} A&=&\frac{3}{2 R_G^2 f^2 (1-f)^2} \nonumber \\
B&=&\frac{R_G^2}{2 f (1-f)} \\ \mchi&=&2\biggl(\chi N- (\chi
N)_s\biggr)+\left(\frac{3}{f^3 (1-f)^3} \right)^{1/2}\nonumber
\end{eqnarray} with $f=N_A/N$ being the volume fraction of the
A-component. In \cite{fr_surf} the third- and fourth-order vertex
functions were assumed to be constant. However, as it was noticed in
\cite{j_rheol,ik_hexag,ik_beta}, it is crucial to keep the
angle-dependence of the fourth-order vertex function in order to
discriminate between the parallel and the perpendicular
orientations. There, the following approximation was made:
\begin{eqnarray}
\label{app1} &&\qquad\qquad\qquad\Gamma_3({\bv q}_1,{\bv q}_2,{\bv
q}_3)=\delta({\bv q}_1+{\bv q}_2+{\bv q}_3)\Gamma_3 \nonumber \\
&&\Gamma_4(\hat{{\bv k}},\hat{{\bv q}},-\hat{{\bv k}},-\hat{{\bv
q}})=\lambda \left(1-\beta (\hat{{\bv k}}\cdot\hat{{\bv q}})^2 \right)
\qquad,\qquad \beta \ll 1
\end{eqnarray} where $\hat{{\bv k}}={\bv k}/k$. In eq.(\ref{app1}) all
the wave-vectors are assumed to have the same length $q_0=\sqrt{A/B}$,
which corresponds to the first unstable mode on the spinodal
\cite{Leibler}. The assumption $\beta\ll1$ was shown to be correct for
almost every architecture of AB block-copolymer molecules
\cite{ik_beta} (for example, for diblocks $\beta\le0.1$). For an
arbitrary star of 4 ${\bv q}$'s one can write to the lowest order in
angles \cite{Brazovskii:1987,klm}
\begin{eqnarray}
\label{app2} &&\qquad\qquad\qquad\qquad\Gamma_4(\hat{{\bv
q}}_1,\hat{{\bv q}}_2,\hat{{\bv q}}_3,\hat{{\bv
q}}_4)=\delta\left(\hat{{\bv q}}_1+\hat{{\bv q}}_2+\hat{{\bv
q}}_3+\hat{{\bv q}}_4\right)\times \nonumber \\ && \qquad\qquad\biggl[
\lambda_0 + \lambda_1\biggl( ({\hv q}_1\cdot{\hv q}_2)({\hv
q}_3\cdot{\hv q}_4) + ({\hv q}_1\cdot{\hv q}_3)({\hv q}_2\cdot{\hv
q}_4) + ({\hv q}_1\cdot{\hv q}_4)({\hv q}_2\cdot{\hv q}_3) \biggr) \\
&&+ \lambda_2\biggl( ({\hv q}_1\cdot{\hv q}_2)^2+({\hv q}_1\cdot{\hv
q}_3)^2+({\hv q}_1\cdot{\hv q}_4)^2+({\hv q}_2\cdot{\hv q}_3)^2+({\hv
q}_2\cdot{\hv q}_4)^2+({\hv q}_3\cdot{\hv q}_4)^2 \biggr)\biggr]
\nonumber \\ &&\qquad\qquad\qquad\qquad\qquad\qquad\qquad
\frac{\lambda_1}{\lambda_0}\text{ , }\frac{\lambda_2}{\lambda_0}\ll1
\nonumber
\end{eqnarray}

Comparison with eq.(\ref{app1}) gives
\begin{equation} \lambda=\lambda_0+\lambda_1+2\lambda_2 \qquad,\qquad
\beta=-\frac{2\lambda_1+4\lambda_2}{\lambda_0+\lambda_1+2\lambda_2}
\end{equation}

Thus, the required real-space representation of the Hamiltonian
${\mathcal H}$ can be written as
\begin{eqnarray}
\label{Ham} &&N {\mathcal H}[\phi]=\int\!\!d{\mathbf {r}} \left[
\frac{B}{2}\biggl({\mathbf \nabla}\phi({\bv
r})\biggr)^2-\frac{1}{2}\mchi\phi({\bv r})^2 +\frac{A}{2}
\int\!\!d{\mathbf r}' {\mathcal G}({\mathbf r}-{\mathbf
r}')\phi({\mathbf r})\phi({\mathbf r}') + \frac{\Gamma_3}{3!}\phi({\bv
r})^3 +\frac{\lambda_0}{4!}\phi({\bv r})^4 \right. \nonumber \\
&&\left. \qquad \qquad + \frac{3 \lambda_1}{4!}\frac{\biggl({\mathbf
\nabla}\phi({\bv r})\cdot{\mathbf \nabla}\phi({\bv
r})\biggr)^2}{q_0^4} + \frac{6 \lambda_2}{4!}\phi({\bv
r})^2\frac{\biggl(\nabla_\alpha \nabla_\beta \phi({\bv
r})\biggr)^2}{q_0^4}-h({\mathbf r})\phi({\mathbf r})\right]+N
{\mathcal H}_s
\end{eqnarray} where
\begin{equation} {\mathcal G}({\mathbf r}-{\mathbf r}')=\int_q
\frac{e^{i {\bv q} ({\bv r}-{\bv r}')}}{q^2}
\end{equation} Here we have added an auxiliary field $h$ which will
help us to construct a thermodynamical potential governing the
dynamics under shear.  Afterwards it will be set to zero.

The Fokker-Planck equation (\ref{FP}) together with
eqns.(\ref{Ham},\ref{const},\ref{Hs}) form a phenomenological set of
equations describing the dynamics of block-copolymer melt under shear
flow in the presence of surfaces. We do not solve these equations
directly, but following \cite{j_rheol} we use the method of Zwanzig
\cite{zwanzig} to derive a system of coupled equations for the first
two cumulants of $P[\phi,t]$
\begin{eqnarray}
\label{cumul} & &c({\bv r})=\langle \phi({\bv r}) \rangle \nonumber \\
S({\bv r}_1,{\bv r}_2)&=&\langle \phi({\bv r}_1)\phi({\bv r}_2)
\rangle- \langle \phi({\bv r}_1) \rangle \langle \phi({\bv r}_2)
\rangle
\end{eqnarray} where $c$ is the average order parameter profile, and
the structure factor $S$ is a measure of the fluctuation' strength. We
introduce a generating functional
\begin{equation} G[\xi,t]=\log \int \!{\mathcal D}\phi
\,\exp\left[\int\!d{\bv r}\,\phi({\bv r})\xi({\bv
r})\right]P[\phi,t]\,\,,
\end{equation} use eq.(\ref{FP}) to derive an equation of motion for
$G[\xi,t]$, and then expand this equation in terms of $\xi$. The two
lowest-order equations read:
\begin{eqnarray}
\label{c} &&\frac{1}{\mu}\frac{\partial c({\bv r})}{\partial
t}=-\frac{D}{\mu}y\frac{\partial c({\bv r})}{\partial x}+B\Delta
c({\bv r})+\mchi c({\bv r})- A \int\!d{\bv r}' \,{\mathcal G}({\bv
r}-{\bv r}')c({\bv r}')-\frac{\Gamma_3}{2}\biggl[c({\bv r})^2+S({\bv
0})\biggr]\nonumber \\ &&\qquad\qquad
-\frac{\lambda_0}{3!}\biggl[c({\bv r})^2+3 S({\bv 0}) \biggr] c({\bv
r})+\frac{\lambda_1}{2q_0^4}\left[\nabla_\alpha \biggl(\nabla_\alpha
c({\bv r}) \left\{{\bv \nabla} c({\bv r})\right\}^2\biggr)
+\tilde{S}_{\alpha \alpha}\Delta c({\bv r})\right] \nonumber \\
&&\qquad -\frac{\lambda_2}{2 q_0^4}\left[c({\bv
r})\biggl(\nabla_\alpha \nabla_\beta c({\bv r})\biggr)^2+c({\bv
r})\tilde S+\nabla_\alpha\nabla_\beta \biggl(c({\bv r})^2
\nabla_\alpha\nabla_\beta c({\bv r})\biggr) +S({\bv 0})\Delta^2 c({\bv
r})\right] \\ &&\qquad
-\frac{2\lambda_1+4\lambda_2}{q_0^4}\biggl(\nabla_\alpha \nabla_\beta
c({\bv r})\biggr)\tilde{S}_{\alpha \beta}+h({\bv r})+\biggl(H_1-a_1
c({\bv r})
\biggr)\left[\delta\left(y+\frac{L}{2}\right)+\delta\left(y-\frac{L}{2}\right)\right]
\nonumber
\end{eqnarray} and
\begin{eqnarray}
\label{S} &&\qquad \qquad \frac{1}{2\mu}\frac{\partial S({\bv r}-{\bv
r}_1)}{\partial t}=\delta({\bv r}-{\bv
r}_1)-\frac{D}{\mu}y\frac{S({\bv r}-{\bv r}_1)}{\partial x}+B \Delta
S({\bv r}-{\bv r}_1)+\mchi S({\bv r}-{\bv r}_1)\nonumber \\
&&-A\int\!d{\bv r}'\,{\mathcal G}({\bv r}-{\bv r}')S({\bv r}'-{\bv
r}_1)-\frac{\Gamma_3}{2}\biggl[S({\bv 0})+2 c({\bv r})S({\bv r}-{\bv
r}_1)\biggr]-\frac{\lambda_0}{2}\biggl[c({\bv r})^2 + S({\bv
0})\biggr] S({\bv r}-{\bv r}_1) \nonumber \\ &&\qquad \qquad
-\frac{\lambda_1+2\lambda_2}{q_0^4}\tilde{S}_{\alpha
\beta}\nabla_\alpha\nabla_\beta S({\bv r}-{\bv
r}_1)+\frac{\lambda_1}{2
q_0^4}\nabla_\alpha\biggl[2\biggl(\nabla_\alpha c({\bv
r})\biggr)\biggl(\nabla c({\bv r})\cdot \nabla S({\bv r}-{\bv r}_1)
\biggr) \nonumber \\ &&\qquad+\biggl(\nabla_\alpha S({\bv r}-{\bv
r}_1)\biggr)\biggl(\nabla c({\bv r})\biggr)^2+\tilde{S}_{\beta \beta}
\nabla_\alpha S({\bv r}-{\bv r}_1) \biggr]-\frac{\lambda_2}{2
q_0^4}\biggl[S({\bv r}-{\bv r}_1)\biggl(\nabla_\alpha \nabla_\beta
c({\bv r})\biggr)^2 \\ &&\qquad \qquad+S({\bv r}-{\bv r}_1)
\tilde{S}+2 c({\bv r})\biggl(\nabla_\alpha \nabla_\beta c({\bv
r})\biggr)\biggl(\nabla_\alpha \nabla_\beta S({\bv r}-{\bv
r}_1)\biggr)+ S({\bv 0})\Delta^2 S({\bv r}-{\bv r}_1) \nonumber \\
&&\qquad \qquad+2\nabla_\alpha \nabla_\beta \biggl(c({\bv r})S({\bv
r}-{\bv r}_1)\nabla_\alpha \nabla_\beta c({\bv r})\biggr) +
\nabla_\alpha \nabla_\beta \biggl(c({\bv r})^2 \nabla_\alpha
\nabla_\beta S({\bv r}-{\bv r}_1)\biggr) \biggr]\nonumber \\ &&\qquad
\qquad\qquad \qquad\qquad-a_1 S({\bv r}-{\bv
r}_1)\left[\delta\left(y+\frac{L}{2}\right)+\delta\left(y-\frac{L}{2}\right)\right]
\nonumber
\end{eqnarray} where
\begin{eqnarray} \tilde{S}_{\alpha\beta}=\nabla'_\alpha \nabla'_\beta
S({\bv r}-{\bv r}')\biggl|_{{\bv r}'={\bv r}} \qquad,\qquad
\tilde{S}=\nabla_\alpha \nabla_\beta \nabla'_\alpha \nabla'_\beta
S({\bv r}-{\bv r}')\biggl|_{{\bv r}'={\bv r}}
\end{eqnarray} Here we have neglected all higher cumulants and made
use of a natural assumption $S({\bv r}_1,{\bv r}_2)=S({\bv r}_1-{\bv
r}_2)$.

Apart from the surface terms, eqns.(\ref{c},\ref{S}) are the
real-space analog of the eqns.(2.25-26) from \cite{j_rheol}. Here the
terms proportional to $S({\bv 0})$ play the role of the fluctuation
integral $\sigma({\hv k})$ from \cite{j_rheol}:
\begin{equation}
\label{sigma} \sigma({\hv k})=\frac{\lambda}{2}\int_q S({\bv
q})\left[1-\beta({\hv k}\cdot{\hv q})^2 \right]
\end{equation}

To keep our model as simple as possible we leave only the linear term
in the surface energy (\ref{Hs}) and put $a_1=0$. Then we set
\begin{eqnarray}
\label{profile} c({\bv r})&=&2 a \cos(q_0 {\bv n}\cdot {\bv
r}+\varphi) \\ h({\bv r})&=&2 h \cos(q_0 {\bv n}\cdot {\bv r}+\varphi)
\nonumber
\end{eqnarray} where $a$ is yet to be determined amplitude, $\bv n$ is
a unit vector perpendicular to the surface of the lamellae and
$\varphi$ is a phase shift which will be chosen to minimize the
surface energy. The auxiliary field $h$ simply follows the behaviour
of $c$.  Fredrickson has shown \cite{fr_surf} that in equilibrium the
presence of the surfaces causes spatial variations of the amplitude
$a$ which decay exponentially away from the surface. Since we are only
interested in the orientation of the lamellar profile (\ref{profile}),
we ignore the spacial dependence of $a$ and set it constant.  With
these simplifications the equation for the Fourier transform $S({\bv
k})$ of $S({\bv r}_1-{\bv r}_2)$ from eq.(\ref{S}) reads
\begin{equation}
\label{SS} \frac{1}{2\mu}\frac{\partial S({\bv k})}{\partial
t}=1+\frac{D}{2\mu}k_x \frac{\partial S({\bv k})}{\partial k_y}-
S_0^{-1}({\bv k})S({\bv k})
\end{equation} where
\begin{eqnarray}
\label{notation} S_0^{-1}({\bv k})&=&r-{\hv k}\cdot \stackrel
{\leftrightarrow}{\bv e} \cdot{\hv k}+B k^2+\frac{A}{k^2}-\mchi_s
\nonumber \\ r-{\hv k}\cdot \stackrel {\leftrightarrow}{\bv e}
\cdot{\hv k}&=&2\biggl(\left(\chi N\right)_s-\chi N \biggr)+\lambda
a^2 \left(1-\beta ({\bv n}\cdot{\hv k})^2 \right)+\sigma({\hv k})
\end{eqnarray} Here we have introduced the same notation as in
\cite{j_rheol,ik_hexag}. In eq.(\ref{notation}) $S_0({\bv k})$ is the
equilibrium structure factor and $r-{\hv k}
\cdot\stackrel{\leftrightarrow}{\bv e}\cdot {\hv k}$ denotes the {\it
renormalized} temperature. Within the fluctuation theory the spinodal
temperature determined from the condition
\begin{equation}
\label{spin_cond} r-{\hv k} \cdot \stackrel {\leftrightarrow}{\bv e}
\cdot {\hv k}\biggr|_{a=0}=0
\end{equation} differs from the mean-field value $\biggl(\chi N
\biggr)_s$. In the case $\beta=0$ such a fluctuation correction was
discussed in \cite{Fredrickson:1987}.  The presence of shear breaks
the rotational symmetry and the spinodal temperature becomes {\it
orientation-dependent}. This gives rise to the $-{\hv k}\cdot
\stackrel {\leftrightarrow}{\bv e} \cdot{\hv k}$ term, with
$e_{ij}\sim\beta$ (see eq.(\ref{app1})).  Here the role of the
angle-dependency in $\Gamma_4$ is especially transparent: if
$\beta=0$, we would not be able to discriminate between different
orientations.

The method of characteristics \cite{van_kampen} gives a formal
solution for the eq.(\ref{SS}):
\begin{equation}
\label{sol_S} S({\bv k},t)=\mu \int_0^t d\tau \exp\left[-\mu
\int_0^\tau ds \, S_0^{-1}(k_x,k_y+\frac{1}{2}D s k_x,k_z) \right]
\end{equation} The steady-state regime is approached as
$t\rightarrow\infty$. The integration in (\ref{sol_S}) can be
performed in the limiting cases $D\rightarrow0$ and
$D\rightarrow\infty$ and will be discussed in the next section.

Now we derive an equation for the amplitude $a$. We substitute the
lamellar profile (\ref{profile}) into eq.(\ref{c}) and perform an
averaging over the lamellar period
\begin{equation} \langle \cdots \rangle =\frac{n_x n_y n_z
q_0^3}{(2\pi)^3}\int_0^{\frac{2\pi}{q_0 n_x}} dx
\int_0^{\frac{2\pi}{q_0 n_y}} dy \int_0^{\frac{2\pi}{q_0 n_z}} dz
\cos\left(q_0 {\bv n}\cdot {\bv r}+\varphi \right) \cdots
\end{equation} Discarding the transverse orientations with $n_x\ne0$
\cite{onuki_kawasaki,cates_milner,j_rheol}, we obtain
\begin{equation}
\label{ae} \frac{1}{\mu}\frac{\partial a}{\partial t}=h-(r-{\bv n}
\cdot \stackrel {\leftrightarrow} {\bv e} \cdot {\bv n})
a+\frac{1}{2}\lambda (1-\beta)a^3+\eta \cos(\varphi)\delta_{n_y^2,1}
\end{equation} where $\eta=\frac{q_0}{\pi}H_1$, and $\delta_{n_y^2,1}$
is the Kronecker delta-symbol which is non-zero only for the parallel
($|n_y|=1$) orientation. Following \cite{j_rheol} we notice that the
equation (\ref{ae}) has a gradient form (with $h=0$):
\begin{equation}
\label{pot_eq} \frac{1}{\mu}\frac{\partial a}{\partial
t}=-\frac{1}{2}\frac{\partial \Phi}{\partial a}
\end{equation} Since the potential $\Phi$ can only decrease with time:
\begin{equation} \frac{\partial \Phi}{\partial t}=\frac{\partial
\Phi}{\partial a}\frac{\partial a}{\partial
t}=-\frac{\mu}{2}\left(\frac{\partial \Phi}{\partial a}\right)^2 <0
\qquad,
\end{equation} the steady-state of the system will be determined by
the minimum of $\Phi$. Now we use the auxiliary field $h$ to construct
$\Phi$. In steady-state $\partial a/\partial t=0$, and $\Phi$ is
obtained by integrating
\begin{equation} h=\frac{1}{2}\frac{\partial \Phi}{\partial a}
\end{equation} Using $h$ from eq.(\ref{ae}), we obtain
\begin{equation}
\label{pot} \Phi=\Phi_0-2\eta a \delta_{n_y^2,1}
\end{equation} where
\begin{equation}
\label{pot0} \Phi_0=-\frac{1}{4}\lambda(1-\beta)a^4+2\int_0^a da'
(r-{\bv n} \cdot \stackrel {\leftrightarrow} {\bv e} \cdot {\bv n})a'
\end{equation} In eq.(\ref{pot}) we have already minimized with
respect to the phase shift $\varphi$, assuming that $a>0$ (the other
terms depend only on even powers of $a$ and are not influenced by this
choice).

The non-trivial dependency of $r-{\bv n} \cdot \stackrel
{\leftrightarrow} {\bv e} \cdot {\bv n}$ on $a$ comes from the term
proportional to $\sigma({\hv k})$ in (\ref{notation}) and the
potential $\Phi$ appears to be dependent on the fluctuation integral
via eq.(\ref{pot0}). Now we are ready to discuss the stable
orientations in different regimes.

\section{Phase transitions} \setcounter{equation}{0}
\subsection*{Crossover from small- to high-shear rate behaviour}

In this subsection we analyze the transition from the parallel to
perpendicular orientation caused by increase of shear rate (the
A-transition in Fig.\ref{fig1}). We start with noticing that at low
shear rates the parallel orientation is the only stable one. Indeed,
as it was shown by Fredrickson \cite{j_rheol}, $\Phi_0$ is minimal for
$n_y^2=1$ in the limit $D\rightarrow0$. The surface term in
eq.(\ref{pot}) also favours the parallel orientation. Thus, our theory
does not modify Fredrickson's prediction for small shear rates.

At high shear rates $D\rightarrow\infty$, the integration in
eq.(\ref{sol_S}) can be performed \cite{onuki_kawasaki}, yielding
\begin{equation}
\label{S_inf} S_{\infty}({\bv k})=c_0 \left(\frac{\mu
q_0^2}{\sqrt{\alpha} D |k_x k_y|} \right)^{2/3}
\end{equation} where
\begin{equation}
c_0=\frac{\Gamma\left(\frac{1}{3}\right)}{(9\pi)^{1/3}} \qquad
\text{and} \qquad \alpha=\frac{q_0^2 B}{\pi}
\end{equation} For the intermediate shear rates $S({\bv k})$ can be
interpolated between $S_0$ and $S_\infty$ \cite{cates_milner}
\begin{equation}
\label{interpol} S({\bv k})=\left[r-{\hv k} \cdot \stackrel
{\leftrightarrow}{\bv e} \cdot {\hv k}+B k^2+\frac{A}{k^2}-\mchi_s+
\frac{1}{c_0}\left(\frac{\sqrt{\alpha} D |k_x k_y|}{\mu q_0^2}
\right)^{2/3} \right]^{-1}
\end{equation} One should realize that the previous equation is an
analytic continuation of the $D\rightarrow\infty$ behaviour to
$D<\infty$ values.  As a result, a small-$D$ behaviour of
eq.(\ref{interpol}) does not correspond to the $D\rightarrow0$
behaviour of eq.(\ref{sol_S}).  On contrary, it describes the $D\sim
O(1)$ region. Since we expect the A-transition to lay in-between the
$D\sim O(1)$ and $D\rightarrow\infty$ regions, we need to calculate
the fluctuation integral $\sigma({\hv k})$ in-between these
regions. This can be done in several steps. First, we use $S({\bv k})$
from eq.(\ref{interpol}) to perform the radial part of the integral in
eq.(\ref{sigma}). This gives
\begin{equation} \sigma({\hv k})\approx\frac{\lambda q_0^2
\sqrt{c_0}}{16\pi^3}\left(\frac{\mu}{D \sqrt{\alpha}}
\right)^{1/3}\int d\Omega \frac{1-\beta\left({\hv k}\cdot{\hv q}
\right)^2 }{|\hat{k}_x
\hat{k}_y|^{1/3}}\biggl[\frac{\pi}{2}+\arctan\left(\frac{q_0\sqrt{c_0}}{|\hat{k}_x
\hat{k}_y|^{1/3}}\left(\frac{\mu}{D \sqrt{\alpha}}\right)^{1/3}\right)
\biggr]
\end{equation} As a next step we expand the integrand for $D\ll1$ and
$D\gg1$ and sum these expressions keeping only the few first
terms. Integration over the orientations of the unit vector $\hv q$
($\int d\Omega\equiv\int_0^\pi d\theta \sin \theta
\int_0^{2\pi}d\phi$) then gives
\begin{eqnarray}
\label{sigma_inter} &&\quad\sigma({\hv k})=\frac{\lambda}{64
\pi^{5/2}}\sqrt{\frac{\alpha}{B^3}}\left\{-4\pi
\left(1-\frac{\beta}{3} \right) +\frac{1}{3 Z^3}\biggl[I_1-\beta
\biggl(I_2 (\hat{k}_x^2+\hat{k}_y^2)+I_3 \hat{k}_z^2 \biggr)
\biggr]\right. \nonumber \\ &&\left.+ 4\pi^2 2^{1/3}\sqrt{3} Z
\biggl[1-\frac{\beta}{7}(2 \hat{k}_x^2+2 \hat{k}_y^2+3 \hat{k}_z^2)
\biggr]+Z^2 \biggl[I_4-\beta \biggl(I_5 (\hat{k}_x^2+\hat{k}_y^2)+I_6
\hat{k}_z^2 \biggr) \biggr] \right\}
\end{eqnarray} where
\begin{eqnarray} &&\qquad \qquad Z=\sqrt{4\pi
c_0}\left[\frac{\sqrt{\alpha}}{\lambda}\frac{D_*}{D}
\right]^{1/3}\qquad,\qquad D_*=\lambda \mu \sqrt{\alpha} \nonumber \\
&&I_1=2\frac{\sqrt{\pi}\gf{5}{6}^2}{\gf{13}{6}} \qquad
I_2=2\frac{\sqrt{\pi}\gf{5}{6}\gf{11}{6}}{\gf{19}{6}} \qquad
I_3=2\frac{\gf{5}{6}^2 \gf{3}{2}}{\gf{19}{6}} \nonumber \\ &&\quad
I_4=\frac{\gf{1}{6}^3}{\sqrt{\pi}} \quad \qquad
I_5=2\frac{\sqrt{\pi}\gf{1}{6}\gf{7}{6}}{\gf{11}{6}} \qquad
I_6=2\frac{\gf{3}{2}\gf{1}{6}^2}{\gf{11}{6}} \nonumber
\end{eqnarray} This procedure is ill-defined from mathematical point
of view. However, any direct analysis of eq.(\ref{sol_S}) is
impossible and we use eq.(\ref{sigma_inter}) for moderate shear rates.

Since eq.(\ref{sigma_inter}) does not depend on the renormalized
temperature, the integration in eq.(\ref{pot0}) is trivial and gives
\begin{equation} \Phi=\left[\tau+\sigma({\bv n})
\right]a^2+\frac{1}{4}\lambda a^4 (1-\beta)-2\eta a \delta_{n_y^2,1}
\end{equation} where
$$ \tau=2\biggl((\chi N)_s-\chi N \biggr)
$$

\begin{figure}
\begin{picture}(0,610)
\put(20,0){\epsfig{file=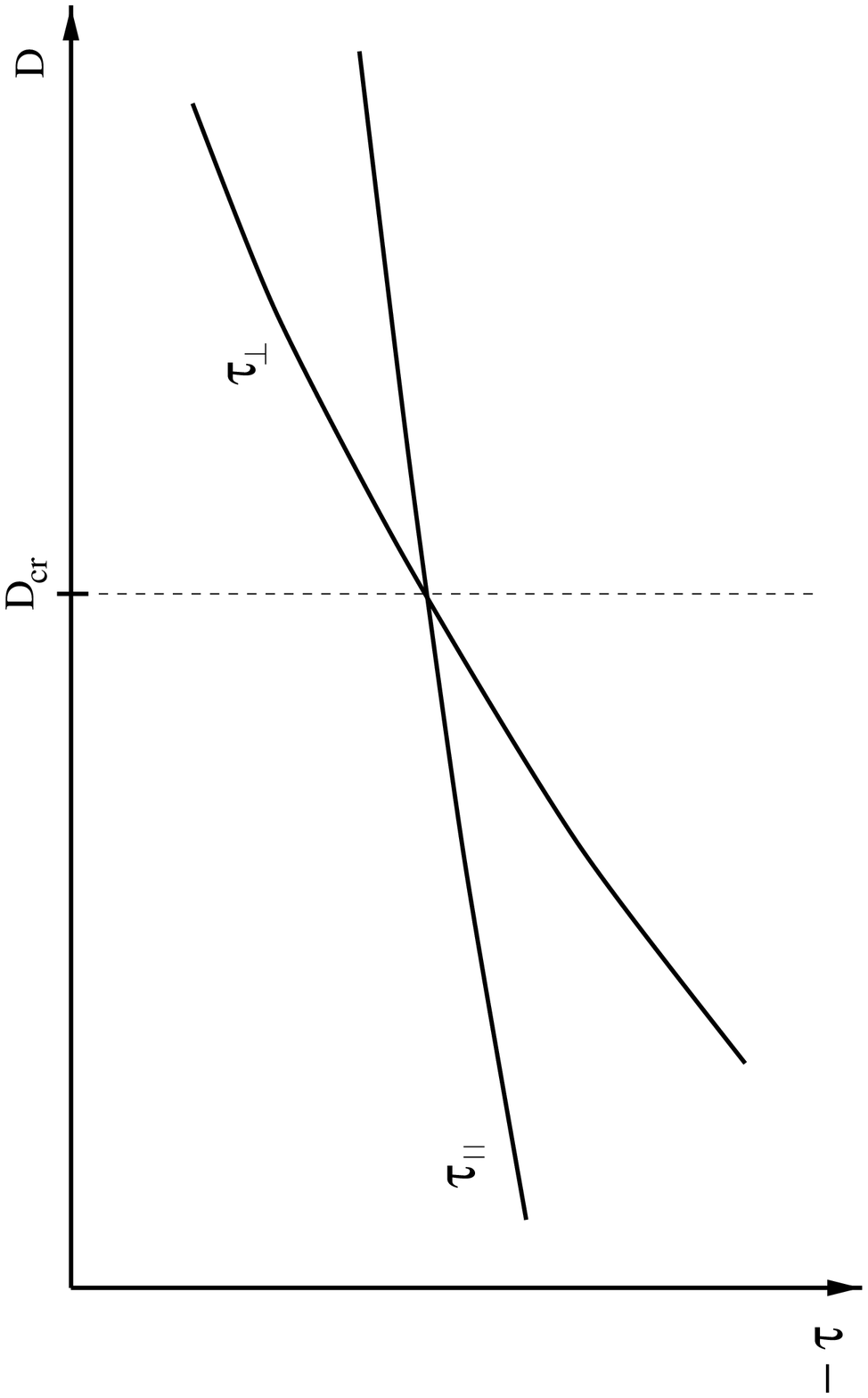,width=13cm}}
\end{picture}
\caption{}
\label{fig3}
\end{figure}

Minimization with respect to $a$ gives to the first order in $\eta$
\begin{equation}
\label{pot_fin} \Phi=-\frac{\left[\tau+\sigma({\bv n})
\right]^2}{\lambda (1-\beta)}-2\eta \sqrt{-2\frac{\tau+\sigma({\bv
n})}{\lambda (1-\beta)}}\delta_{n_y^2,1}
\end{equation} The order-disorder transition (ODT) occurs when $\Phi$
becomes negative. The corresponding transition temperature is
\begin{equation} \tau_s({\bv n})=-\sigma({\bv n})
\end{equation} which coincides with eq.(\ref{spin_cond}). The
orientation with the lowest $\sigma$ will appear immediately below the
ODT temperature.  The crossover (the A-transition) is then located at
such a value of $D$ that $\sigma_{\parallel}-\sigma_{\perp}$ changes
its sign. From eq.(\ref{sigma_inter}) this point is given by
\begin{equation}
\label{eq_Z} \sigma_{\parallel}-\sigma_{\perp}\sim\frac{1}{3
Z^2}(I_3-I_2)+\frac{4\pi^2}{7}2^{1/3}\sqrt{3}Z+Z^2(I_6-I_5)=0
\end{equation} and is found to be
\begin{equation}
\label{Dcr} D_{cr}\approx 2.64 \mu \alpha
\end{equation} From eq.(\ref{eq_Z}) it also follows that
$\tau_s^{\parallel}>\tau_s^{\perp}$ for $D<D_{cr}$, which fits the
small-shear behaviour discussed in the beginning of this
subsection. When $D>D_{cr}$, the perpendicular orientation first
appear below the spinodal. This crossover is depicted in
Fig.\ref{fig3}.

As it was noticed before \cite{cates_milner,j_rheol}, the mean-field
behaviour is restored in the limit $D\rightarrow\infty$. On the other
hand, the small-$D$ region is dominated by fluctuations. Thus,
$D_{cr}$ can be interpreted as a position of a crossover from the
fluctuation to mean-field behaviour. The scaling properties of
$D_{cr}$ follow from eq.(\ref{Dcr}) and are determined by the Onsager
coefficient $\mu$. Using the results of
\cite{kawasaki_sekimoto,Binder:1983,Fredrickson:1986,muthu_shear}
($\mu\equiv q_0^2 \lambda(q_0)/N$ with $\lambda$ from
\cite{Binder:1983,Fredrickson:1986}) we obtain
\begin{equation} D_{cr}\sim N^{-3}
\end{equation} which shows that the fluctuation region disappears in
the limit $N\rightarrow\infty$. In equilibrium the same conclusion was
drawn in \cite{Leibler,Fredrickson:1987}.

Finally, we emphasize that the results of this subsection are
independent of the surface interaction. The same results can be
obtained within the Fredrickson theory \cite{j_rheol} ($\eta=0$).

\subsection*{B- and C-transitions} In the previous subsection we have
discussed the order-disorder transitions. Now we consider lower
temperatures and look for transitions between different orientations
in the high-shear limit. The corresponding free energies are given by
eq.(\ref{pot_fin})
\begin{eqnarray} &&\Phi_\parallel=-\frac{\left(\tau+\sigma_\parallel
\right)^2}{\lambda (1-\beta)}-2\eta
\sqrt{-2\frac{\tau+\sigma_\parallel}{\lambda (1-\beta)}} \\
&&\qquad\qquad\Phi_\perp=-\frac{\left(\tau+\sigma_\perp
\right)^2}{\lambda (1-\beta)}\nonumber
\end{eqnarray} where $\sigma({\bv n})$ is given by its high-shear
limit of eq.(\ref{sigma_inter})
\begin{equation} \sigma({\bv n})=\frac{\left(\alpha \lambda
\right)^{2/3}}{B^{3/2}}\left(\frac{D_*}{D}
\right)^{1/3}\frac{2^{1/3}\sqrt{3}}{8}\sqrt{c_0}\biggl[1-\frac{\beta}{7}(2
n_y^2+3 n_z^2)\biggr]
\end{equation}

To the leading order in $D_*/D$, the transition from the perpendicular
to parallel orientation occurs at temperatures which are the roots of
the equation $\Phi_\parallel=\Phi_\perp$
\begin{equation}
\label{taus}
\tau_1=-\sigma_\parallel-\frac{(\sigma_\parallel-\sigma_\perp)^4}{8
\eta^2 \lambda (1-\beta)} \qquad \tau_2=-\frac{2 \eta^2 \lambda
(1-\beta)}{(\sigma_\parallel-\sigma_\perp)^2} \qquad \text{if} \quad
\eta^2 \lambda (1-\beta)>(\sigma_\parallel-\sigma_\perp)^3
\end{equation} where
\begin{equation}
\sigma_\parallel-\sigma_\perp=\beta\frac{2^{1/3}\sqrt{3
c_0}}{56}\frac{\left(\alpha \lambda
\right)^{2/3}}{B^{3/2}}\left(\frac{D_*}{D} \right)^{1/3}
\end{equation} There $\tau_1$ corresponds to the
$\perp\rightarrow\parallel$ transition, while $\tau_2$ - to the
reverse one. Now we summarize our results in a phase diagram.

\section{Discussion of the phase diagram and Conclusion}
\setcounter{equation}{0}

\begin{figure}
\begin{picture}(0,610)
\put(-50,0){\epsfig{file=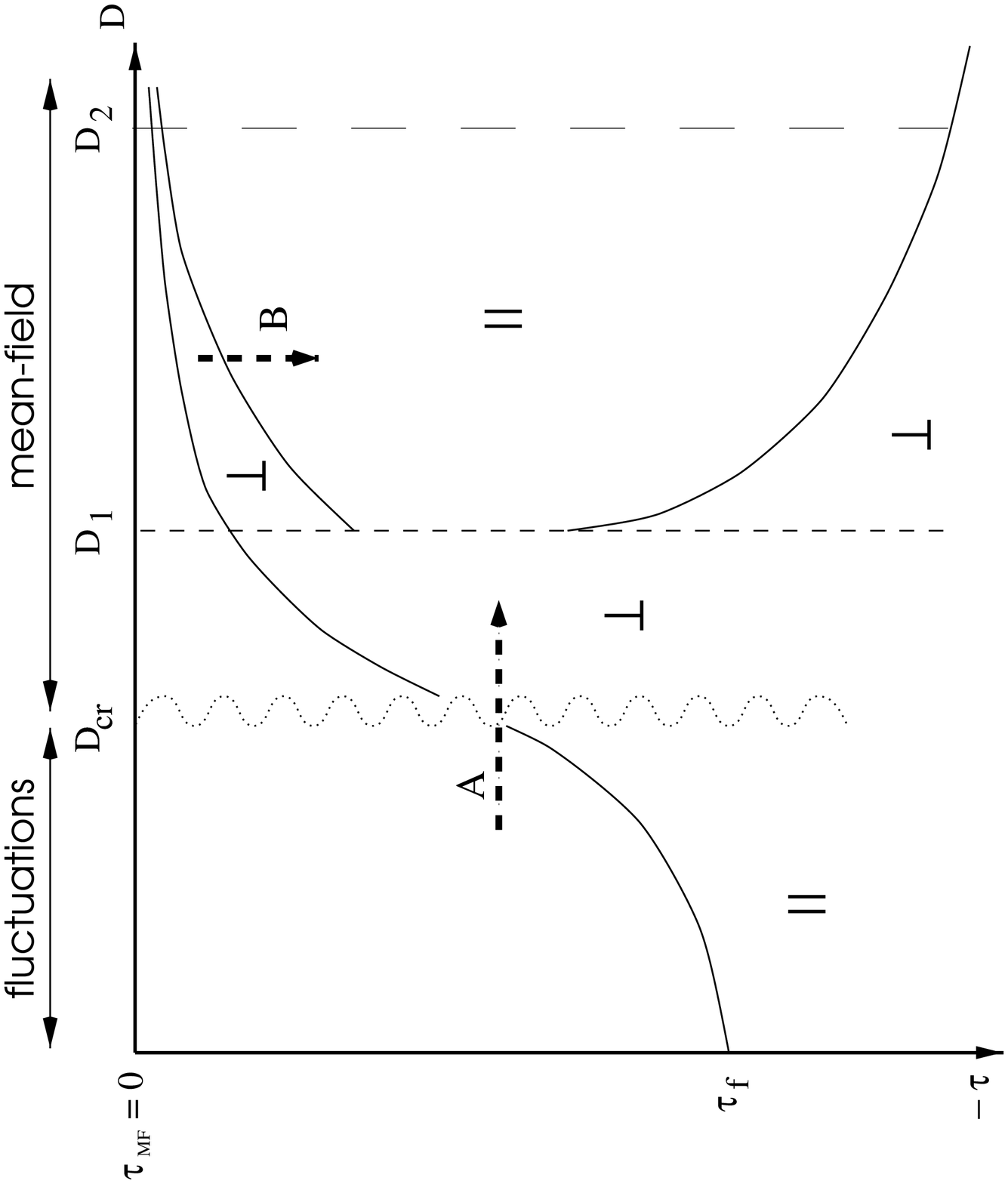,width=18cm}}
\end{picture}
\caption{}
\label{fig4}
\end{figure}

In this work we incorporated some real-system properties into the
previously developed theory of the orientational phase transitions
under shear flow \cite{cates_milner,j_rheol,ik_hexag}. In particular,
we considered the influence of the shear-cell boundaries in the
gradient direction on the orientation of the lamellar phase. In
equilibrium the lamellae are known to orient parallel with respect to
the boundaries \cite{fr_surf,binder_films}. Under shear the tendency
to orient parallel to the surfaces competes with the orientation
favoured by the flow which appears as a result of the coupling between
the flow velocity field and the order-parameter fluctuations
\cite{cates_milner,j_rheol}. The interplay between these two factors
produces the non-trivial phase diagram shown in Fig.\ref{fig4}.

At low shear rates the parallel orientation is preferred by both the
shear and surface terms in eq.(\ref{pot}). Therefore it is the only
stable orientation in that part of the phase diagram. When shear rate
reaches the value $D_{cr}$ given by eq.(\ref{Dcr}), the perpendicular
orientation becomes stable immediately below the ODT temperature. We
associate this change in orientation with the crossover from the
fluctuation-dominated behaviour to the mean-field one. Indeed, at very
small shear rates the equilibrium fluctuation spectrum is only
slightly modified by the flow, while at high shear rates the flow
strongly suppresses fluctuations and restores the mean-field
behaviour. Therefore, there is a crossover point and the corresponding
change of orientation.

At high shear rates and away from the spinodal, the surface influence
starts to play an important role. In the narrow region between
$D_{cr}$ and $D_1$, estimated from the condition in eq.(\ref{taus})
$$ D_1=2\frac{(3 c_0)^{3/2}}{56^3}\frac{\beta}{1-\beta}D_*
\frac{\lambda \alpha^2}{\eta^2 B^{9/2}}
$$ the influence of shear is still very strong and is capable of
stabilizing the perpendicular orientation at all temperatures.  The
size of this region is very small due to the scaling $D_1\sim
N^{-13/2}$. When $D>D_1$ there appears a region where the parallel
orientation is stable. It takes over the perpendicular one at
$\tau=\tau_1$ and looses its stability again at $\tau=\tau_2$ given by
eq.(\ref{taus}). This region grows as the shear rate increases, and in
the limit $D=\infty$ the parallel orientation occupies the whole range
of temperatures $(0,-\infty)$. This coincides with the predictions of
the equilibrium mean-field theory \cite{fr_surf,binder_films}.  We
therefore argue that there is no sharp C-transition as shown in
Fig.\ref{fig1}. Since the region between the spinodal and the parallel
phase shrinks with an increase of the shear rate, there always be some
value $D_2$ such that for $D>D_2$ the size of this region will be
smaller than the resolution of the experimental device. This value
$D_2$ can be misinterpreted as a position of an additional transition.

An important feature of our theory is that it is able to reproduce the
B-transition without additional assumptions. In the previous theory
\cite{j_rheol}, Fredrickson had to take into account the difference in
viscosities of the pure components in order to reproduce the
B-transition. Namely, he put $\eta[\phi]=\eta_0+\eta_1 \phi$ which can
be considered as a Taylor expansion of the viscosity $\eta[\phi]$. As
a result, in high-shear limit the size of the stability region for the
perpendicular phase is of order of $(\eta_0/\eta_1)^2$ and grows as
$D\rightarrow\infty$. While depicting the main physics, this approach
has internal problems since the derivative $\eta_1$ is not a
well-defined object and therefore the whole theory depends on a
phenomenological parameter which is difficult to estimate. Moreover,
Fredrickson's theory does not predict the C-transition. Our theory is
free from these problems. It, however, predicts the
$\parallel\rightarrow\perp$ transition at very low temperatures in
high-shear limit which was not observed. There could be several
explanations of this prediction. First, this transition occurs at very
low temperatures ($\tau_2$) where the weak-segregation theory does not
work. Second, this transition might be an artifact of the $O(\eta)$
expansion. Finally, this transition can be removed if we use
Fredrickson's argument about the viscosity dependence on the order
parameter. It stabilizes the parallel orientation at low temperatures.

It is possible that the absolute value of the surface interaction is
small. However, since it acts as a {\it symmetry-breaking} factor, its
influence is very important \cite{balsara94,laurer}. This statement
can be checked experimentally. We have shown that the positions of the
transition lines in the high-shear limit are dependent on the strength
of the surface-copolymer interaction $\eta\sim(\chi
N)_{cop-surf}$. Therefore, the phase diagram of a particular copolymer
system depends on the material of the shear-cell walls. A systematic
study of this dependency will provide arguments for or against our
theory.

In this work we have considered the influence of the walls in the
gradient direction. We also want to comment on the role of the
boundaries in the other shear directions (flow and
vorticity). Formally, these walls will also induce alignment parallel
to themselves. However, the flow profile near those walls is no longer
a simple triangular one and we expect this disordered flow to destroy
their orientational tendency. Moreover, the distance between those
surfaces is normally much larger than between the walls in the
gradient direction and their influence is thus weaker. Therefore we
neglected them in our work.

Finally, we want to discuss briefly possible modifications and
extensions of the developed theory. A very interesting problem is to
calculate the alterations of the density profile in a confined system
under shear. This can be achieved by restoring the position dependency
in the amplitude $a$ and deriving the corresponding amplitude equation
from eq.(\ref{c}). In the absence of shear this problem was solved in
\cite{fr_surf}. Another possibility is to use our formalism for other
external fields rather than interactions with surfaces. A good example
is an electric field which is coupled to the square of the order
parameter \cite{helf_electr,onuki_electr}. With some modifications
eq.(\ref{c},\ref{S}) can be a starting point for the corresponding
theory. Importance of such a theory for a system in electric field and
simultaneously under shear was outlined in \cite{onuki_electr}.

\section*{Acknowledgments} The authors want to express their gratitude
to Eugenij Polushkin for helpful discussions of the experimental
details and Berk Hess for careful reading of the manuscript.

\renewcommand{\theequation}{\thesection.\arabic{equation}}

\appendix
\section*{Influence of thermal fluctuations}

Here we want to emphasize the role of fluctuations in our theory. We
are going to show that if the Fokker-Planck equation (\ref{FP}) is
replaced by a deterministic one, the theory will not be able to
discriminate between different orientations. What follows should not
be considered as a proof but, more likely, as an illustration that has
general features.

Let us consider a Langevin equation equivalent to the Fokker-Planck
equation (\ref{FP}). If we now remove the noise term, it reads
\begin{equation}
\label{lang} \frac{\partial \phi}{\partial t}+{\bv
v}\cdot\nabla\phi=\tilde{\mu}\nabla^2\frac{\delta {\mathcal
H}_L}{\delta \phi}
\end{equation} where an Onsager mobility $\tilde{\mu}$ differs from
$\mu$ in eq.(\ref{FP}) and ${\mathcal H}_L$ is the Leibler Hamiltonian
(eq.(\ref{Ham})) without ${\mathcal H}_s$ and $h=0$. A similar
equation was considered in \cite{drolet}. There the authors used a
Hamiltonian with $\lambda_1=\lambda_2=0$ and showed that in
steady-state the theory predicts both orientations to be equally
stable at all shear rates. This is not surprising since their theory
does not contain fluctuations and the angular dependence of the
fourth-order vertex function $\Gamma_4$ -- the two ingredients that
were argued to be crucial in explaining the reorientation phenomena
\cite{j_rheol,ik_hexag,ik_beta}.

In order to separate these two effects we keep the angular dependence
in $\Gamma_4$ ($\lambda_1$, $\lambda_2\ne 0$), but use the
deterministic equation (\ref{lang}). We follow the approach of
\cite{drolet} and derive an amplitude equation from eq.(\ref{lang})
assuming a single plane-wave density profile. The solution of this
equation is
\begin{equation} A(t)=\biggl[\frac{P(t)}{A(0)^2}+3 P(t)\int_0^t d\tau
f(\tau) P^{-1}(\tau) \biggr]^{-1/2}
\end{equation} and
\begin{eqnarray} f(t)=1+4 Q^4
\left(\frac{\lambda_1+2\lambda_2}{\lambda_0} \right) \qquad , \qquad
P(t)=\exp \left[-2\int_0^t d\tau \sigma(\tau)\right] \nonumber \\
\sigma(t)=(1+\epsilon) Q^2-Q^4-\frac{1}{4} \qquad , \qquad
Q^2=q_x^2+(q_y-D t q_x)^2+q_z^2 \nonumber
\end{eqnarray} where $A$ is the amplitude,
$\epsilon=(\mchi-\mchi_s)/\mchi_s$, and the amplitude and the units of
length and time are scaled with $\sqrt{\mchi_s/\lambda_0}$,
$\sqrt{B/\mchi_s}$ and $B/(\tilde{\mu}\mchi_s^2)$, respectively. We
see that the equation becomes symmetric with respect to the
interchange $q_y\leftrightarrow q_z$. Thus, even in the presence of
the angular dependence in $\Gamma_4$ it is impossible to distinguish
between the parallel and perpendicular orientations starting from
eq.(\ref{lang}).

We believe that a theory without fluctuations of the order parameter
is not capable of describing the reorientational transitions. There
are phenomena where fluctuations only modify a deterministic behaviour
\cite{Hamley00,ren01,vlim99}. However, the orientational behaviour
under shear flow does not belong to this class of phenomena. It can
only occur in the presence of thermal fluctuations.

\newpage

\section*{FIGURE CAPTIONS}

\vspace{1cm}

Fig.1. Schematic phase diagram ``temperature vs. shear rate'' as a
compilation of the theoretical predictions by Fredrickson and the
experimentally observed C-transition. The order-disorder transition
line behaves as $\tau\sim D^2$ for small shear rates $D$ and $\tau\sim
D^{-1/3}$ for $D\rightarrow\infty$. The B-transition line levels off
in Fredrickson's theory.

\vspace{1cm}

Fig.2. Orientations of the lamellar phase in a simple shear flow. The
axes of the coordinate system correspond to the shear geometry: $x$ is
the flow direction (${\bv v}$), $y$ is the gradient direction ($\nabla
v_x$) and $z$ is the vorticity direction ($\nabla\times{\bv v}$). In
the parallel orientation the normal to the lamellar layers is oriented
parallel to the gradient direction, in the perpendicular - to the
vorticity direction. The walls in $y$ direction interact with the melt
and prefer one of the components.

\vspace{1cm}

Fig.3. Schematic behaviour of the spinodal temperatures for the
parallel and perpendicular orientations in the vicinity of the
crossover point.

\vspace{1cm}

Fig.4. Phase diagram for the lamellar phase under steady simple shear
flow as predicted in this work.

\begin{thebibliography}{10}

\bibitem{hamley:book} I. Hamley, {\em The Physics of Block Copolymers}
(Oxford University Press, Oxford, New York, Tokyo, 1998).

\bibitem{Leibler} L. Leibler, Macromolecules {\bf 13}, 1602 (1980).

\bibitem{koppi92} K.~A. Koppi {\it et~al.}, J. Phys. France II {\bf
2}, 1941 (1992).

\bibitem{koppi93} K.~A. Koppi, M. Tirrell, and F.~S. Bates,
Phys. Rev. Lett. {\bf 70}, 1449 (1993).

\bibitem{tepe} T. Tepe {\it et~al.}, Macromolecules {\bf 28}, 3008
(1995).

\bibitem{walter98} G. Schmidt, W. Richtering, P. Lindner, and
P. Alexandridis, Macromolecules {\bf 31}, 2293 (1998).

\bibitem{cates_milner} M.~E. Cates and S.~T. Milner,
Phys. Rev. Lett. {\bf 62}, 1856 (1989).

\bibitem{j_rheol} G.~H. Fredrickson, J. Rheol. {\bf 38}, 1045 (1994).

\bibitem{ik_hexag} A.~N. Morozov, A.~V. Zvelindovsky, and
J.~G. E.~M. Fraaije, Phys. Rev. E {\bf 61}, 4125 (2000).

\bibitem{zhang95} Y. Zhang, U. Wiesner, and H.~W. Spiess,
Macromolecules {\bf 28}, 778 (1995).

\bibitem{patel95} S.~S. Patel, R.~G. Larson, K.~I. Winey, and
H. Watanabe, Macromolecules {\bf 28}, 4313 (1995).

\bibitem{fr_surf} G.~H. Fredrickson, Macromolecules {\bf 20}, 2535
(1987).

\bibitem{binder_films} K. Binder, Adv. Pol. Sci. {\bf 138}, 1 (1998).

\bibitem{balsara94} N.~P. Balsara {\it et~al.}, Macromolecules {\bf
27}, 2566 (1994).

\bibitem{laurer} J.~H. Laurer, B.~S. Pinheiro, D.~L. Polis, and
K.~I. Winey, Macromolecules {\bf 32}, 4999 (1999).

\bibitem{freed92} H. Tang and K.~F. Freed, J. Chem. Phys. {\bf 97},
4496 (1992).

\bibitem{binder94} M. Kikuchi and K. Binder, J. Chem. Phys. {\bf 101},
3367 (1994).

\bibitem{chakrabarti} G. Brown and A. Chakrabarti, J. Chem. Phys. {\bf
101}, 3310 (1994).

\bibitem{chakrabarti94} G. Brown and A. Chakrabarti,
J. Chem. Phys. {\bf 102}, 1440 (1994).

\bibitem{binder97} K. Binder, H.~L. Frisch, and S. Stepanow,
J. Phys. II France {\bf 7}, 1353 (1997).

\bibitem{matsen97} M.~W. Matsen, J. Chem. Phys. {\bf 106}, 7781
(1997).

\bibitem{agur_film} H.~P. Huinink, J.~C.~M. Brokken-Zijp, M.~A. van
Dijk, and G.~J.~A. Sevink, J.  Chem. Phys. {\bf 112}, 2452 (2000).

\bibitem{agur_dynamics} G.~J.~A. Sevink {\it et~al.},
J. Chem. Phys. {\bf 110}, 2250 (1999).

\bibitem{onuki_kawasaki} A. Onuki and K. Kawasaki, Ann. Phys. {\bf
121}, 456 (1979).

\bibitem{Ohta:1986} T. Ohta and K. Kawasaki, Macromolecules {\bf 19},
2621 (1986).

\bibitem{Cruz:1986} M.~O. de~la Cruz and I.~C. Sanchez, Macromolecules
{\bf 19}, 2501 (1986).

\bibitem{ik_beta} A. Morozov and J. Fraaije, 2001, submitted to
Macromolecules.

\bibitem{Brazovskii:1987} S.~A. Brazovskii, I.~E. Dzyaloshinskii, and
A.~R. Muratov, Sov. Phys. - JETP {\bf 66}, 625 (1987).

\bibitem{klm} E.~I. Kats, V.~V. Lebedev, and A.~R. Muratov,
Phys. Rep. {\bf 228}, 1 (1993).

\bibitem{zwanzig} R. Zwanzig, in {\em {L}ecture {N}otes in {P}hysics,
Vol. 132: {S}ystems {F}ar from {E}quilibrium}, edited by L. Garrido
(Springer, New York, 1980), pp.\ 198--225.

\bibitem{Fredrickson:1987} G.~H. Fredrickson and E. Helfand,
J. Chem. Phys. {\bf 87}, 697 (1987).

\bibitem{van_kampen} N.~G. van Kampen, {\em Stochastic processes in
physics and chemistry} (North-Holland, Amsterdam, Oxford, New York,
Tokyo, 1981).

\bibitem{kawasaki_sekimoto} K. Kawasaki and K. Sekimoto,
Macromolecules {\bf 22}, 3063 (1989).

\bibitem{Binder:1983} K. Binder, J. Chem. Phys. {\bf 79}, 6387 (1983).

\bibitem{Fredrickson:1986} G.~H. Fredrickson, J. Chem. Phys. {\bf 85},
5306 (1986).

\bibitem{muthu_shear} C.-Y. Huang and M. Muthukumar,
J. Chem. Phys. {\bf 107}, 5561 (1997).

\bibitem{helf_electr} K. Amundson, E. Helfand, X. Quan, and
S.~D. Smith, Macromolecules {\bf 26}, 2698 (1993).

\bibitem{onuki_electr} A. Onuki and J. Fukuda, Macromolecules {\bf
28}, 8788 (1995).

\bibitem{drolet} F. Drolet, P. Chen, and J. Vi{\~n}als, Macromolecules
{\bf 32}, 8603 (1999).

\bibitem{Hamley00} I.~W. Hamley, Macromol. Theory Simul. {\bf 9}, 363
(2000).

\bibitem{ren01} S.~R. Ren and I.~W. Hamley, Macromolecules {\bf 34},
116 (2001).

\bibitem{vlim99} B.~A.~C. van Vlimmeren {\it et~al.}, Macromolecules
{\bf 32}, 646 (1999).

\end{thebibliography}
\end{document}